\documentclass{PoS}

\usepackage{amsmath}
\usepackage{mathtools}

\newcommand{\pt}{$p_T$}
\newcommand{\et}{$E_T$}
\newcommand{\met}{$E_T^{miss}$}

\title{Performance of the CMS Level-1 Trigger}

\ShortTitle{Performance of the CMS Level-1 Trigger}

\author{\speaker{J. Brooke}\\
       {\rm On behalf of the CMS Collaboration}\\
       University of Bristol\\
       E-mail: \email{jim.brooke@cern.ch}}

\abstract{The first level trigger of the CMS experiment is comprised of custom electronics that process data from the electromagnetic and hadron calorimeters and three technologies of muon detectors in order to select the most interesting events from LHC collisions, such as those consistent with the production and decay of the Higgs boson. The rate of events selected by this Level-1 trigger must be reduced from the beam crossing frequency to no more than 100~kHz further processing can occur, a major challenge since the LHC instantaneous luminosity has increased by six orders of magnitude since the  start of operations to more than $6 \times 10^{33} {\rm cm^{-2} s^{-1}}$ today. The performance of the Level-1 trigger, in terms of rates and efficiencies of the main objects and trigger algorithms, as measured from LHC proton collisions at 7 and 8 TeV center-of-mass energies is presented here.}

\FullConference{36th International Conference on High Energy Physics,\\
		July 4-11, 2012\\
		Melbourne, Australia}

\begin{document}

\section{Introduction}

One of the central challenges to experimental physics at a hadron collider is the trigger problem. Searches for rare processes such as a Higgs decaying to two photons, which has a predicted cross-section around 50~fb, demand high instantaneous luminosity of the collider, simply to collect a sizable sample of signal events.  During the 2012 run, the CERN Large Hadron Collider (LHC) has operated up to $L=8 \times 10^{33} {\rm cm^{2}s^{-1}}$, corresponding to a proton-proton collision rate of around $5 \times 10^8$~Hz. Given the high segmentation of the LHC detectors, approaching 100 million channels, this corresponds to an enormous volume of data at the detector front-ends, and the selection of signal events must start online. The challenge of the trigger system is therefore to maintain high acceptance for events of interest, at the same time as high rejection of QCD jet events that account for most of the rate. In the case of the search for the Higgs boson at the LHC, events of interest are characterised by relatively low energy leptons and photons, requiring trigger thresholds in the tens of GeV. The Compact Muon Solenoid (CMS) trigger system has maintained low thresholds and high efficiency for these important triggers throughout the 2012 run, enabling a wide range of physics studies, including the discovery of a new boson with a mass around 125 GeV~\cite{CMSHiggs}. The performance of the CMS Level-1 Trigger during 2012 is the subject of this paper. The performance of the CMS Higher Level Trigger is described elsewhere in these proceedings~\cite{HLT}.

The central feature of the CMS apparatus is a superconducting solenoid of 6~m internal diameter, providing a magnetic field of 3.8~T. Within the superconducting solenoid volume are a silicon pixel and strip tracker, a lead tungstate crystal electromagnetic calorimeter (ECAL), and a brass/scintillator hadron calorimeter (HCAL). The ECAL has an energy resolution of better than 0.5\% for unconverted photons with transverse energies above 100~GeV. The HCAL, when combined with the ECAL, measures jets with a resolution $\Delta E/E \approx 100\% / \sqrt{E} \oplus 5\%$. In the central barrel, the HCAL cells have widths of 0.087 in pseudorapidity ($\eta$) and 0.087 in azimuth ($\phi$). In the $\eta$-$\phi$ plane, and for $|\eta|< 1.48$, the HCAL cells map on to $5 \times 5$ ECAL crystals arrays to form calorimeter towers projecting radially outwards from close to the nominal interaction point. At larger values of $|\eta|$, the size of the towers increases and the matching ECAL arrays contain fewer crystals. Within each tower, the energy deposits in ECAL and HCAL cells are summed to define the calorimeter tower energies, subsequently used to provide the energies and directions of hadronic jets. Extensive forward calorimetry complements the coverage provided by the barrel and endcap detectors. Muons are measured in the pseudorapidity range $|\eta|< 2.4$, with detection planes made using three technologies: drift tubes, cathode strip chambers, and resistive plate chambers. Matching muons to tracks measured in the silicon tracker results in a transverse momentum resolution between 1 and 5\%, for \pt values up to 1~TeV. The first level (L1) of the CMS trigger system, composed of custom hardware processors, uses information from the calorimeters and muon detectors to select the most interesting events in a fixed time interval of less than 4~$\mu$s, of which 1~$\mu$s is available for data processsing. The High Level Trigger (HLT) processor farm further decreases the event rate from around 100~kHz to around 300~Hz, before data storage. A more detailed description can be found in Ref.~\cite{CMSPaper}.

In the L1 Trigger, candidate muons are identified in each of the DT, CSC and RPC detectors separately.  In the DT and CSC detectors, track stubs are identified at the chamber/sector level, and forwarded to separate track-finders for each sub-detector (DTTF and CSCTF). Track stubs are shared between these systems to ensure full coverage of the barrel-endcap transition region. The RPC identifies muon candidates directly from hits in pattern comparator logic (RPC PAC).  The identified candidates from all three systems are sent to the Global Muon Trigger (GMT), where they are combined and the four best muon candidates in barrel and endcap are forwarded to the Global Trigger. The ECAL and HCAL calorimeters calculate energy deposits in trigger towers ($0.087 \eta \times 0.087 \phi$ in the barrel) that are sent to the Regional Calorimeter Trigger (RCT), where e/$\gamma$ candidates are identified and energy is further summed into regions ($0.35 \eta \times 0.35 \phi$ in the barrel).  The e/$\gamma$ candidates and region sums are sent to the Global Calorimeter Trigger (GCT) where the e/$\gamma$ candidates are sorted and jets and energy sums are calculated.  The GCT sends four isolated and four non-isolated e/$\gamma$ candidates to the Global Trigger (GT), along with four jets in each of the following categories: central, forward and tau.  The GCT also sends total and missing $E_T$ sums, and total and missing $E_T$ sum over jets ($H_T$), to the GT. The GT is programmed with a menu of up to 128 trigger algorithms that make requirements on the candidates received from the GCT and GMT.  As well as simple thresholds on $E_T$ or $p_T$, the GT can require combinations of objects with requirements on their relative position.

Commissioning of the L1 Trigger with the detectors started in 2007. A large sample of cosmic ray muons was collected in 2008, and used to gain an initial understanding of the trigger performance before the start of LHC collisions~\cite{Chatrchyan:2009ab}. Further commissioning with LHC beams commenced in 2009, before the physics program started in earnest during 2010. As the LHC instantaneous luminosity increased over the course of the 2010-2012 runs, the GT trigger menu evolved to cope with increasing trigger rates.  Table~\ref{menu} shows the main algorithms, in terms of rate, used in the trigger menu for LHC operations at $L=6.6 \times 10^{33} {\rm cm^{2}s^{-1}}$. At this luminosity, the peak L1 trigger rate is~90kHz with a total 3\% deadtime.

\begin{table}[th]
  \caption{Rates of selected trigger algorithms at $L=6.6 \times 10^{33} {\rm cm^{2}s^{-1}}$.}\label{menu}
  	\begin{center}
    	\begin{tabular}{|c|c|c|}\hline
      		Trigger 						& Threshold (GeV) 	& Rate (kHz) \\ \hline
			Single $\mu$ ($\eta < 2.1$)		& 14				& 7		\\
			Double $\mu$ ($\eta < 2.4$)		& 10, 0				& 6		\\
			Single e/$\gamma$				& 20				& 13  	\\
			Double e/$\gamma$				& 13, 7				& 8		\\
			e/$\gamma$ + $\mu$				& 12, 3.5			& 3		\\
			$\mu$ + e/$\gamma$				& 12, 7				& 1.5	\\
			Single jet						& 128				& 1.5	\\
			Quad jet						& 36				& 5		\\
			$H_{T}$							& 150				& 5		\\
			$E_{T}^{miss}$					& 40				& 8		\\ \hline
    \end{tabular}
 \end{center}
\end{table}

\section{Muon Triggers}

\begin{figure}
	[ht] 
	\begin{center}
		\resizebox{0.49\linewidth}{0.39\linewidth}{\includegraphics{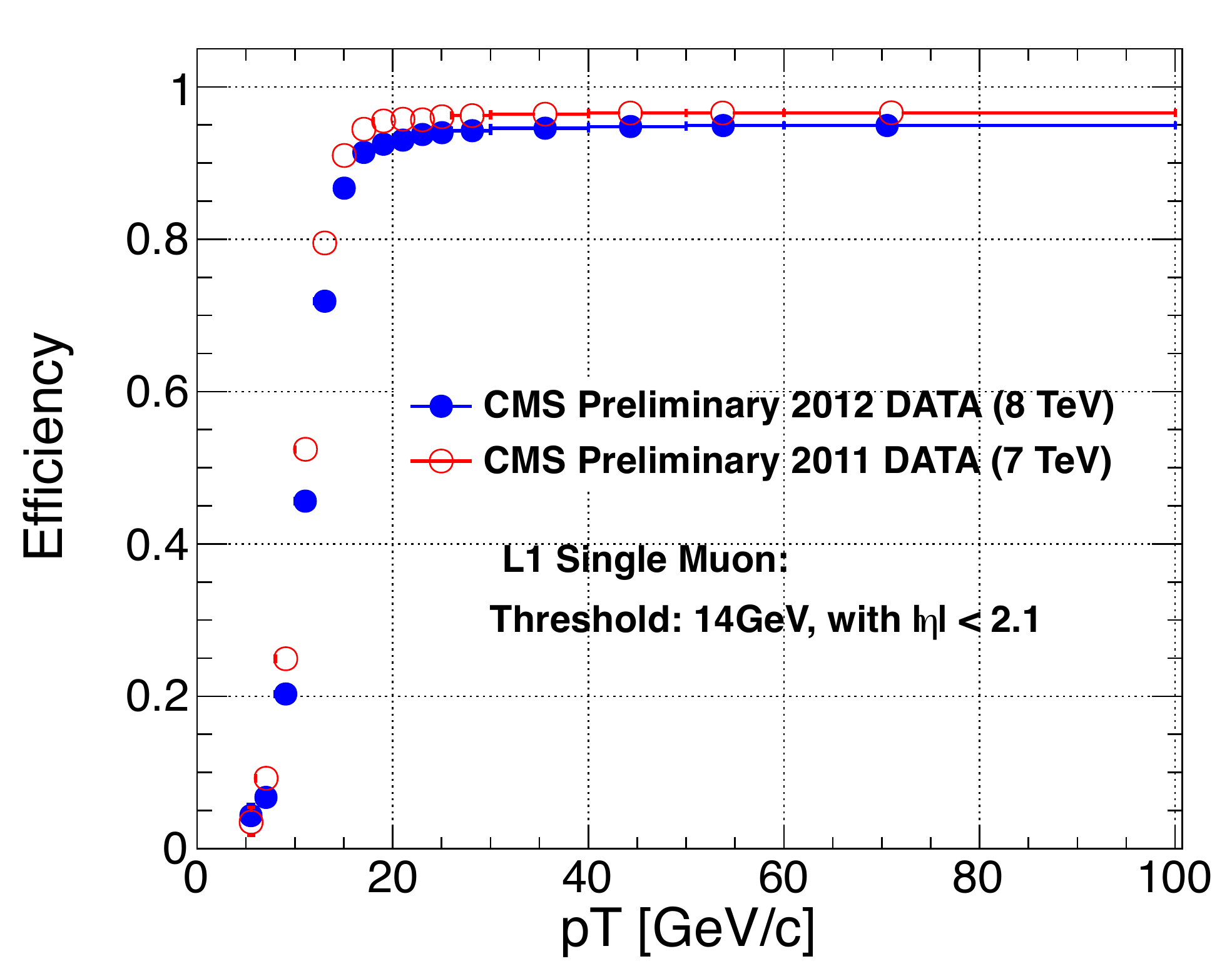}}
		\resizebox{0.49\linewidth}{0.39\linewidth}{\includegraphics{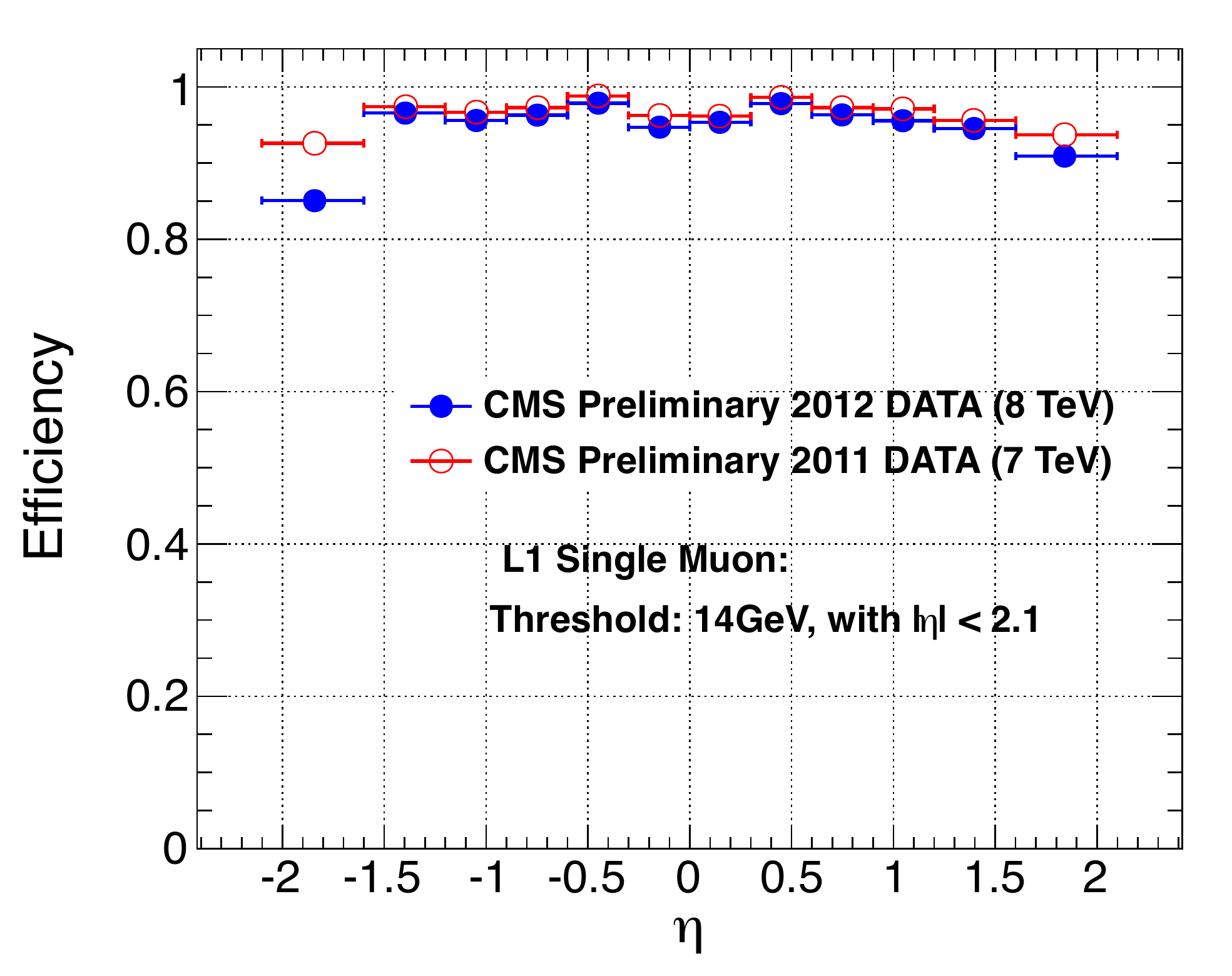}}	
		\caption{{\it Left}: Efficiency of the 14 GeV single muon trigger as a function of offline muon \pt, for $\eta<2.1$. {\it Right}: Efficiency of the 14 GeV single muon trigger as a function of offline muon $\eta$. The rate of this trigger was reduced by around 50\% between 2011 and 2012, at very little cost in efficiency, as is shown.}\label{fig:muon} 
	\end{center}
\end{figure}

The efficiency of the 14 GeV single muon trigger is shown in Figure~\ref{fig:muon} as a function of the offline $p_T$ and $\eta$ of the muon.  After the 2011 run, improvements in $p_T$ assignment were made in the CSCTF, and the use of $p_T$ in merging candidates in the GMT was optimised, reducing the rate over the full acceptance by 50\%.  This was achieved with negligible effect on the efficiency, as shown in Figure~\ref{fig:muon}.

\section{E/$\gamma$ Triggers}

\begin{figure}
	[ht] 
	\begin{center}
		\resizebox{0.41\linewidth}{0.39\linewidth}{\includegraphics{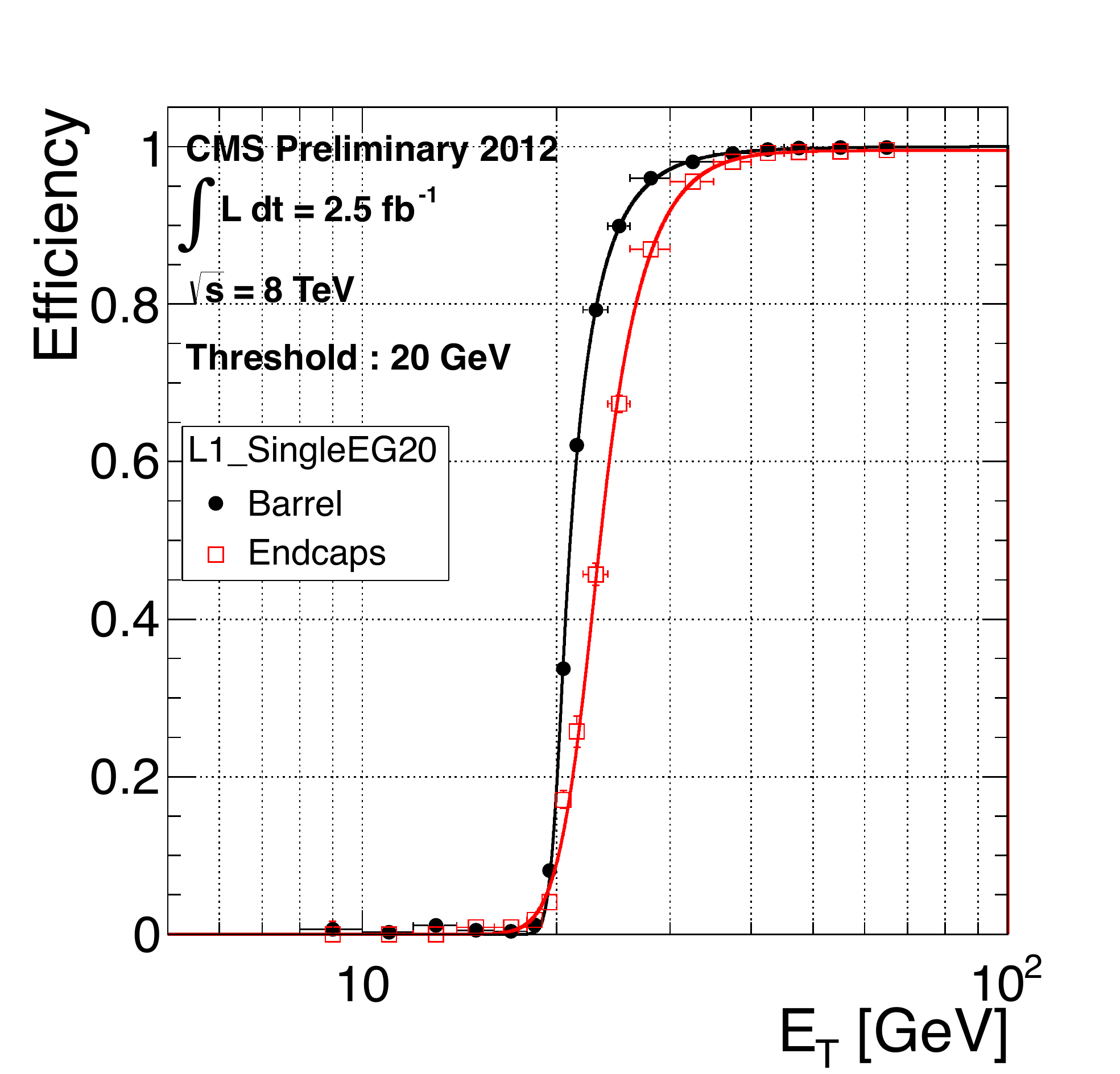}}
		\resizebox{0.58\linewidth}{0.38\linewidth}{\includegraphics{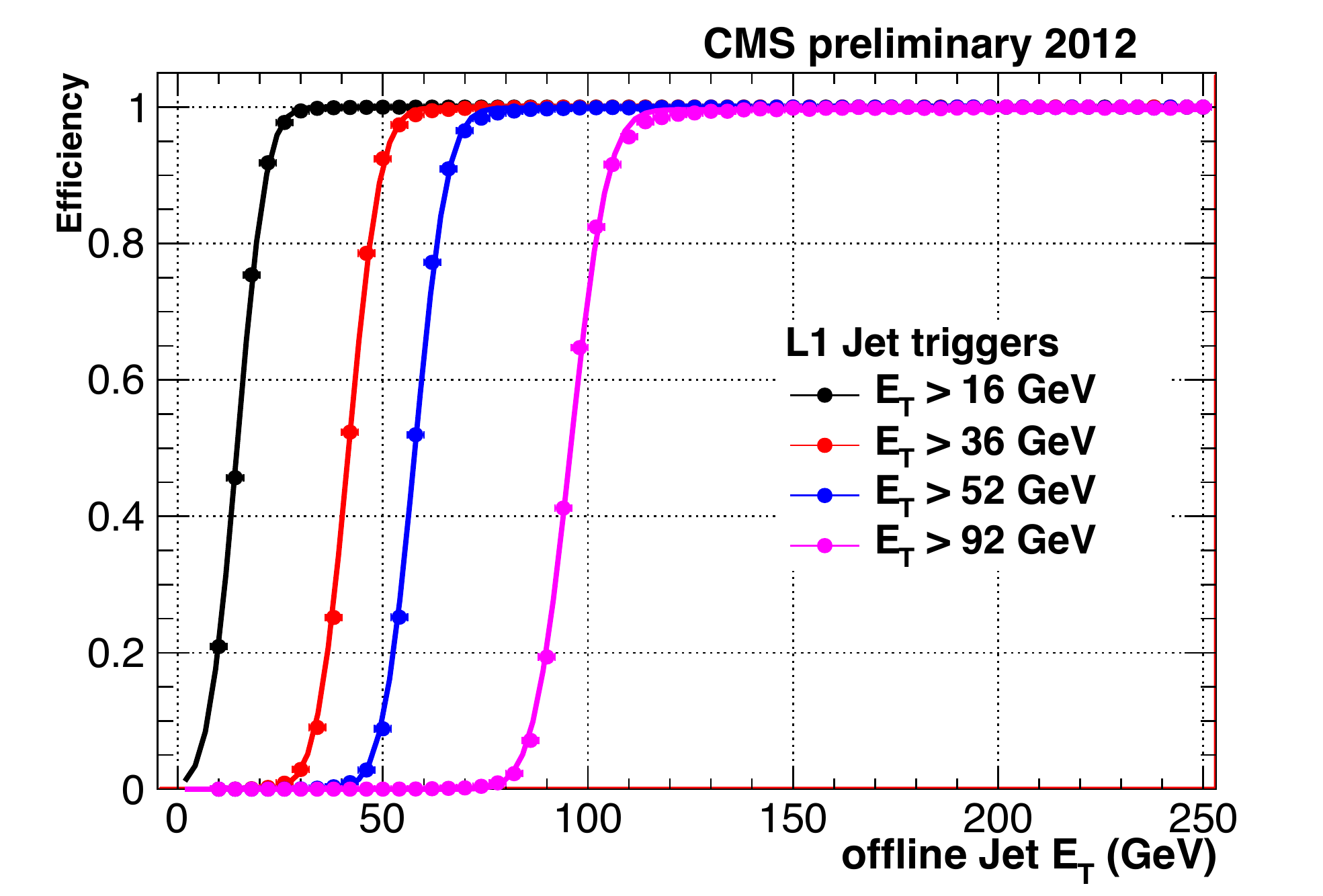}}
		\caption{{\it Left}: Efficiency of the 20 GeV e/$\gamma$ trigger as a function of offline \et, shown separately for barrel and endcap. {\it Right}: jet trigger efficiency as a function of offline jet \et, for several online thresholds.}\label{fig:egJet} 
	\end{center}
\end{figure}

The efficiency of the 20 GeV single electron trigger is shown in Figure~\ref{fig:egJet} (Left), as a function of the offline electron $p_T$.  Transparency corrections for the ECAL endcap crystals were applied in 2012, improving the steepness of the efficiency turn-on relative to that in 2011~\cite{Daci:2011zz}.

\section{Jet and Energy Sum Triggers}

\begin{figure}
	[ht] 
	\begin{center}
		\resizebox{0.49\linewidth}{0.33\linewidth}{\includegraphics{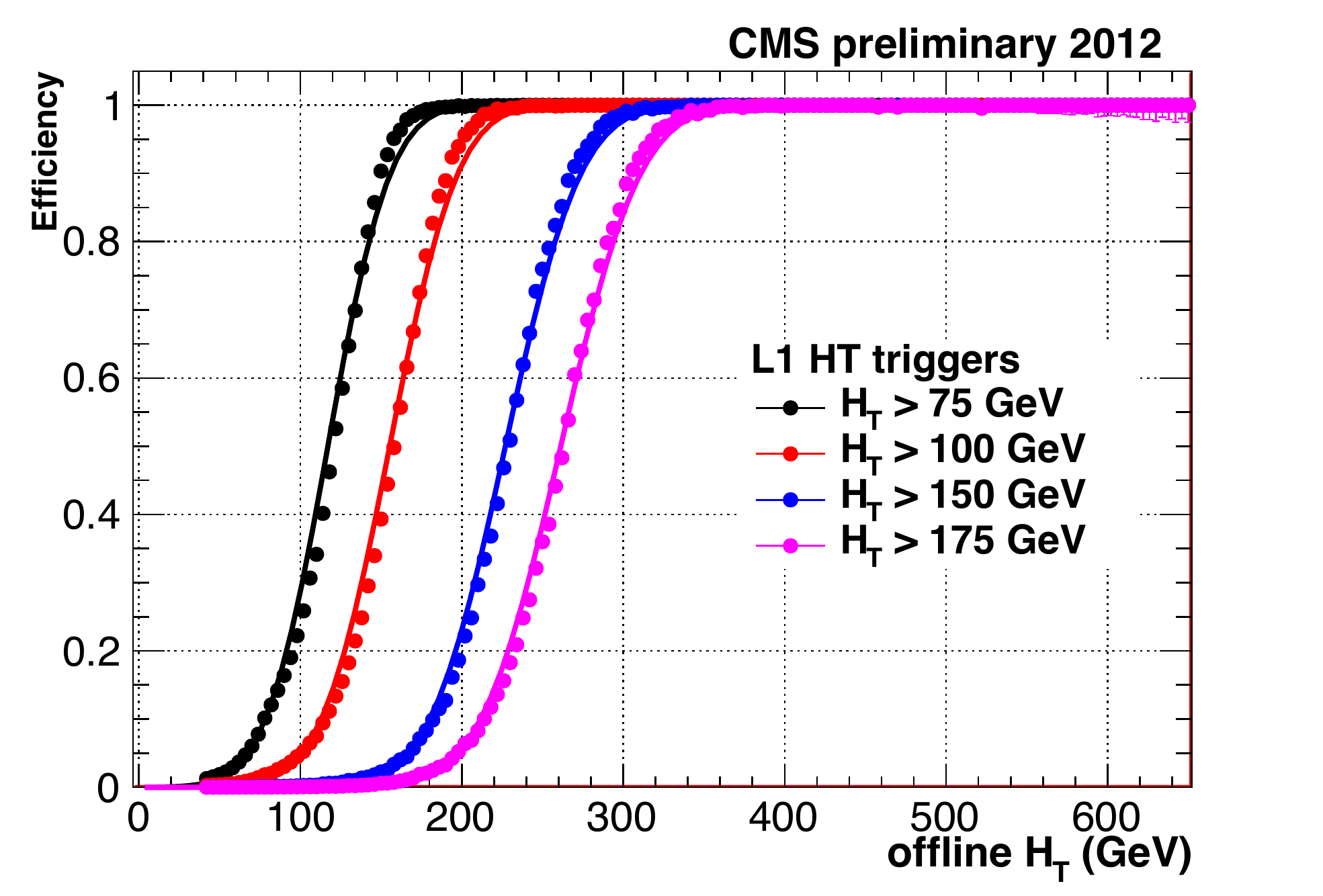}}
		\resizebox{0.49\linewidth}{0.33\linewidth}{\includegraphics{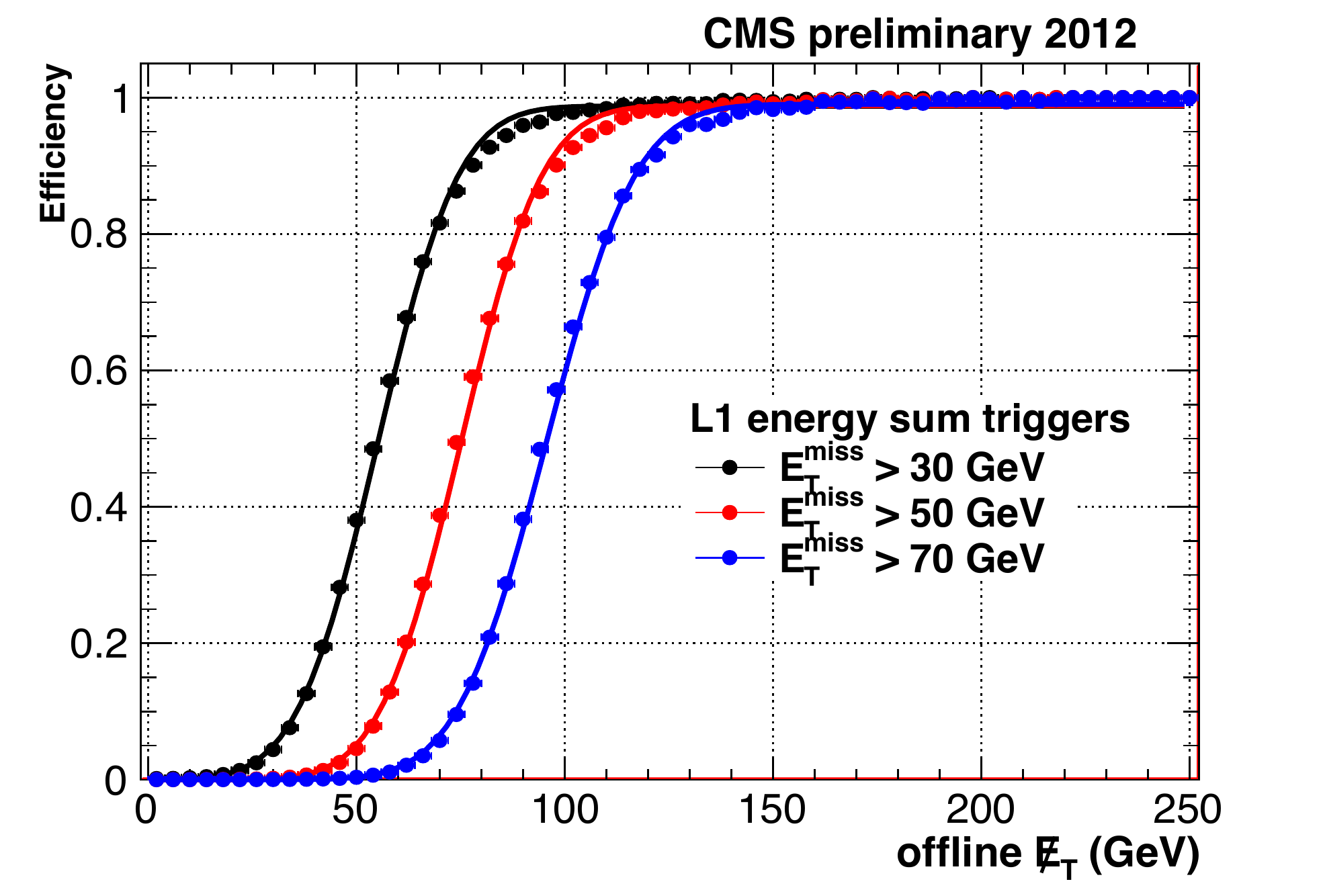}}
		\caption{{\it Left}: $H_T$ trigger efficiency as a function of offline $H_T$, for several online thresholds.  {\it Right}: $E_T^{miss}$ trigger efficiency as a function of offline \met, for several online thresholds.}\label{fig:esums} 
	\end{center}
\end{figure}

\begin{figure}
	[ht] 
	\begin{center}
		\resizebox{0.49\linewidth}{0.33\linewidth}{\includegraphics{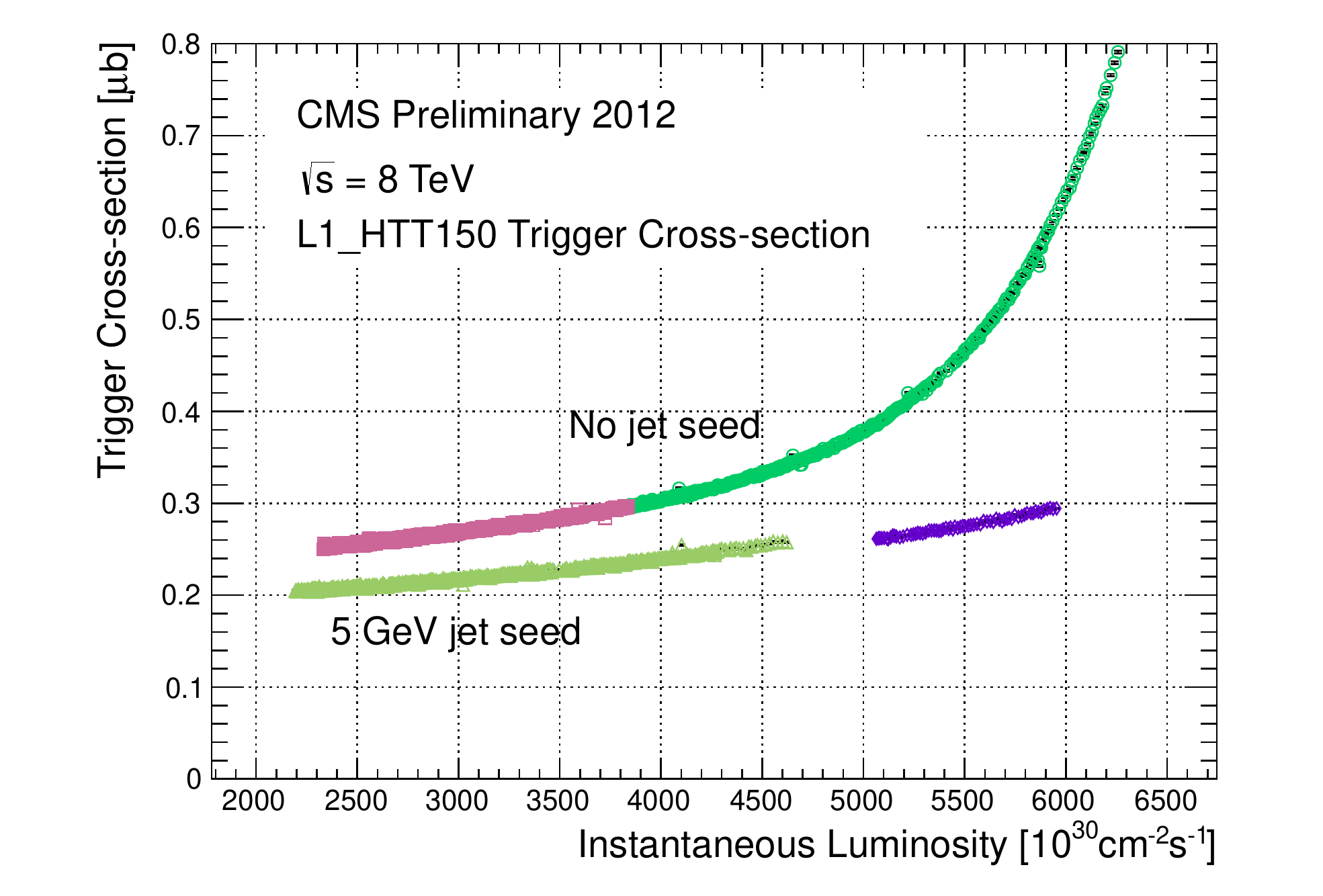}}
		\caption{Cross-section of the 150 GeV $H_T$ trigger, as a function of instantaneous luminosity before and after applying a 5 GeV threshold to the jet algorithm seed region (see text for details). } \label{fig:htt}
	\end{center}
\end{figure}

Jets are identified using a $3 \times 3$ sliding window algorithm running over the region sums.
The efficiency of the single jet trigger, for several thresholds, is shown in Figure~\ref{fig:egJet} (Right) as a function of the offline $p_T$. Multi-jet final states are triggered using $H_T$, the $E_T$ sum over jets with $|\eta|<3$ and $E_T>10$~GeV.  The efficiency of the $H_T$ trigger is shown for a variety of thresholds in Figure~\ref{fig:esums} (Left) as a function of the offline quantity. The increases in LHC luminosity achieved during 2012 resulted in high numbers of simultaneous proton-proton collisions per bunch crossing. The affect on the $H_T$ trigger can be seen from the upper trace in Figure~\ref{fig:htt}, which shows the cross-section of the 150 GeV $H_T$ trigger as a function of the instantaneous luminosity for a single fill taken before the June 2012 LHC technical stop.  At the start of the fill, the large number of simultaneous pp collisions results in a higher average energy density, and hence large numbers of low $E_T$ jets identified by the L1 algorithm.  To counteract this, a 5 GeV threshold was applied to the central region of a jet, reducing the rate of low $E_T$ jets. The lower trace in Figure~\ref{fig:htt} shows the trigger cross-section for a fill after the jet seed threshold was applied, reducing the dependence of the $H_T$ cross-section on instantaneous luminosity.

\section{Summary}

The CMS L1 trigger has delivered events for physics analysis throughout the first phase of LHC operations from 2009 to 2012, enabling the broad CMS physics program, and including discovery of a new boson with a mass around 125~GeV. The trigger has been routinely operated at rates up to 100~kHz, while maintaining low deadtime, around 3\%.  Improvements in the muon trigger logic have allowed the single muon trigger to operate with a threshold of 14~GeV at $L=6 \times 10^{33} {\rm cm^{2}s^{-1}}$, and improvements since the ICHEP conference have enabled a further reduction in threshold to 12~GeV. The single e/$\gamma$ threshold has been maintained at 20~GeV for this instantaneous luminosity. The very high number of simultaneous proton-proton interactions has presented a challenge for hadronic triggers, although this has been mitigated by applying a 5~GeV threshold to the central region of the L1 jet window. The L1 trigger is expected to maintain excellent performance at instantaneous luminosity up to $L=8 \times 10^{33} {\rm cm^{2}s^{-1}}$ until the end of the 2012 run. For operations at higher instantaneous luminosity, an extensive program of upgrades to the L1 trigger are planned, starting during the 2013-4 long LHC shutdown~\cite{Upgrade}.

\end{document}